\documentclass[11pt]{article}
\usepackage{cite}
\usepackage{amsmath, amssymb}
\usepackage{graphicx}
\usepackage{authblk}
\usepackage{blindtext}

\title{Non-hermitian bosonization}
\author{V.K. Sazonov}
\affil{Institute of Physics, Department of Theoretical Physics, University of Graz, Universit\"atsplatz 5, A-8010 Graz, Austria}

\begin{document}

\maketitle

\begin{abstract}
A method for the bosonization of complex actions is presented. Together with the convergent perturbation theory
it provides a conceptually new way for bypassing fermion sign problems.
\end{abstract}

\section{Introduction}
Bosonization of fermionic field theories is one of the most interesting problems of physics. Initially it
was considered within the framework of $(1 + 1)$-dimensional models \cite{Coleman, Polyakov1, Polyakov2, Witten, GamboaSaravi, Furuya, Naon},
where there is no spin and therefore bosons and fermions are similar to each other. The bosonization in $(1 + 1)$-dimensions is a powerful tool
in quantum field theory and condensed matter physics. Its extension to higher dimensions is important for the understanding of the
underlying relations between bosonic and fermionic degrees of freedom, as well as for developing new computational methods.
Several significant results were obtained for the $(2+1)$-dimensional case \cite{Kovner1994, Semenoff1988, Luscher1989, Marino1991, Huerta1993}.
The most general approaches valid for any dimension $d$ were suggested by L\"uscher \cite{Luscher1994} and Slavnov \cite{Slavnov1996, Slavnov1996a, Slavnov1999, Slavnov2000}.
Both of these methods rely on the utilization of auxiliary $(d+1)$-dimensional bosonic matter fields
\footnote{In the L\"uscher multi-boson approach the sum over the bosonic flavors in the action is equivalent to an extra spatial dimension.}.
However, the application of these techniques is restricted only to real (non-complex) actions with an even number of fermionic flavors.
The bosonization of complex actions was an open problem for the decades.
Here we construct a bosonization for complex actions with an even number of flavors. 
Therefore, we establish the possibility to describe fermions not only in vacuum, but also in dense matter as bosonic degrees of freedom.
In particular, our procedure gives a recipe for the bosonization of QCD at finite chemical potentials.

\section{Hermitization of the fermion determinant}
For definiteness we consider the lattice discretization of the path integral. 
The generalization to the continuum case is given at the end of Section \ref{sec:Slavnov}.

Consider the partition sum of two flavors of fermions $\bar{\psi}_{i, x},\,\psi_{i, x}$, $i = 1,2$ interacting with the gauge field $U$
\begin{equation}
  {\cal Z} = \prod_{i = 1,2;\newline x = 1..N}\int [d\bar{\psi}_{i, x}] [d\psi_{i, x}] \int [dU_x] e^{-S_G + \sum_{x, y, i} \bar{\psi}_{x, i} K_{x, y} \psi_{y, i}}\,,
\end{equation}
where $N$ is the $d$-dimensional lattice volume. Extension of the following steps to other even numbers of flavors is trivial.
$K_{x,y}$ is the kernel of the fermion action.
We represent it as $K_{x,y} = A_{x,y} + H_{x,y}$, where $A_{x,y}$ is anti-hermitian and $H_{x,y}$ is hermitian.
Inserting unity with an auxiliary integration and the delta function we change the term $\sum_{y, i} \bar{\psi}_{i, x} H_{x, y} \psi_{i, y}$ 
to a new bosonic variable $f_x$
\begin{eqnarray}
\nonumber
  {\cal Z} = \prod_{i = 1,2;\newline x = 1..N}\int [d\bar{\psi}_{i, x}] [d\psi_{i, x}] \int [dU_x] \\
 \int_{-\infty}^{\infty} [df_x]\,
  e^{-S_G + \sum_{x}\big(\big[\sum_{y, i} \bar{\psi}_{i, x} A_{x, y} \psi_{i, y}\big] + f_{x}\big)} 
  \delta \big(f_{x} - \sum_{y, i} \bar{\psi}_{i, x} H_{x, y} \psi_{i, y}\big)\,.
\label{herm1}
\end{eqnarray}
Representing the delta function as
\begin{equation}
\nonumber
  \delta(a - b) = \frac{1}{2 \pi}\lim_{\lambda \rightarrow 0} \int_{-\infty}^\infty dh\, e^{i h (a - b) - \lambda (h^2 + a^2)}\,,
\label{deltarep}
\end{equation}
we rewrite partition sum as
\begin{eqnarray}
\nonumber
{\cal Z} = \frac{1}{(2 \pi)^N}\lim_{\lambda \rightarrow 0}\prod_{i = 1,2;\newline x = 1..N} \int [d\bar{\psi}_{i, x}] [d\psi_{i, x}] \int [dU_x] \int_{-\infty}^{\infty} [df_x] \\
\int_{-\infty}^{\infty} [dh_x]\,
e^{-S_G + \sum_{x}\big(\big[\sum_{y, i} \bar{\psi}_{i, x} (A_{x, y} - i h_n H_{x, y}) \psi_{i, y}\big] +  f_{x} + i h_x f_{x} - \lambda (h_x^2 + f_x^2)\big)}\,.
\label{herm2}
\end{eqnarray}
Integration over the fermion fields gives the squared determinant of the anti-hermitian operator 
\begin{equation}
\det(A_{x, y} - i h_x H_{x, y})^2 \equiv \det((i B)^2)_{x, y} = (-1)^N \det (B^2)_{x,y}\,,
\label{defB}
\end{equation}
where $B_{x,y} = -i (A_{x, y} - i h_x H_{x, y})$ is hermitian. The factor $(-1)^N$ is not significant, 
since it cancels in all vacuum expectation values of observables.

The transformations (\ref{herm1}), (\ref{herm2}) and definition (\ref{defB}) map the non-hermitian part of the 
fermion determinant to the auxiliary bosonic fields. 
Therefore, the initial problem reduces to the bosonization of the determinant of a hermitian matrix.

\section{Application to the Slavnov bosonization}
\label{sec:Slavnov}
The utilization of the hermitization for the Slavnov bosonization procedure is straightforward. Following \cite{Slavnov1996},
we write
\begin{eqnarray}
\nonumber
\det (B^2)_{x,y} = \lim_{\alpha \rightarrow 0, b \rightarrow 0} \int [d\phi_{k, x}] [d\phi^\dagger_{k, x}] \int [d\chi_{k, x}] [d\chi^\dagger_{k, x}]\\ 
e^{a^4 b\sum_x \sum_{k = -n+1}^{n}\sum_x \big[\alpha\frac{\phi^\dagger_{k+1, x} - \phi^\dagger_{k, x}}{b}\phi_{k, x} 
- \sum_y [ \phi^\dagger_{k, x} (B^2)_{x, y} \phi_{k, y}]
- \frac{i}{\sqrt{L}}(\phi^\dagger_{k, x}\chi_x + \chi^\dagger_x \phi_{k, x})\big]}\,.
\end{eqnarray}
Here $\phi_{k, x}$, $\phi^\dagger_{k, x}$ are $d+1$ dimensional bosonic fields, carrying the same indices as the initial fermions.
The fields $\chi_{k, x}$, $\chi^\dagger_{k, x}$ are $d$-dimensional and implement the non-local constraint
\begin{equation}
  \sum_{k = -n + 1}^n \phi_k(x) = \sum_{k = -n + 1}^n \phi_k^\dagger(x) = 0\,.
\end{equation}
Then, the partition function is given by the expression
\begin{eqnarray}
\nonumber
{\cal Z} = \frac{1}{(2\pi)^N}\lim_{\lambda \rightarrow 0} \lim_{\alpha \rightarrow 0, b \rightarrow 0} \int [d\phi_{k, x}] [d\phi^\dagger_{k, x}] \int [dU_x] \int [d\chi_{x}] [d\chi^\dagger_{x}]\\
\nonumber
\int_{-\infty}^{\infty} [df_x] \int_{-\infty}^{\infty} [dh_x]\,
e^{-S_G +  \sum_x\big[ f_{x} + i h_x f_{x} - \lambda (h_x^2 + f_x^2) \big]}\\
e^{a^4 b\sum_x \sum_{k = -n+1}^{n}\big[\alpha\frac{\phi^\dagger_{k+1, x} - \phi^\dagger_{k, x}}{b}\phi_{k, x} 
+ \sum_y [\phi_{k, x} (A_{x, y} - i h_x H_{x, y})^2\phi^\dagger_{k, y}]
- \frac{i}{\sqrt{L}}(\phi^\dagger_{k, x}\chi_x + \chi^\dagger_x \phi_{k, x})\big]}\,.
\label{SlavnZ}
\end{eqnarray}
Here the order of limits is crucial for the convergence of the integrals and the limit $\lambda\rightarrow 0$ must be taken last.
The continuum limit of equation (\ref{SlavnZ}) is
\begin{eqnarray}
\nonumber
{\cal Z} = \frac{1}{(2\pi)^N}\lim_{\lambda \rightarrow 0} \lim_{\alpha \rightarrow 0} \int D\phi(x, \tau) D\phi^\dagger(x, \tau) \int DA_\mu(x) \int D\chi(x) D\chi^\dagger(x)\\
\nonumber
\int Df(x) \int Dh(x)\,
e^{-S_G[A_\mu] +  \int\,d^dx\big[ f(x) + i h(x) f(x) - \lambda (h^2(x) + f^2(x)) \big]}\\
\nonumber
\cdot
e^{\int d\tau\,\int d^dx \big[\alpha \partial_\tau \phi^\dagger(x, \tau)\phi(x, \tau) + \big(\int\, d^dy\, \phi(x, \tau) [A(x, y) - i h(x) H(x, y)]^2\phi^\dagger(y, \tau)\big)\big]}\\
\cdot
e^{-i\int_0^1 d\tau\,\int d^dx\, (\phi^\dagger(x, \tau)\chi(x) + \chi^\dagger(x) \phi(x, \tau))}\,,
\label{SlavnZC}
\end{eqnarray}
where $S_G[A_\mu]$ denotes the continuum gauge action.

\section{Application to the L\"uscher bosonization}
In the L\"uscher approach the matrix $(B^2)_{x,y}$ must be bounded, but this is not the case. To bound $(B^2)_{x,y}$, 
we borrow the damping factor $e^{ - \lambda / 2 \sum_x  h_x^2}$ from the delta function and represent it as 
a determinant of a diagonal matrix
\begin{equation}
  (T^2)_{x,y} = \text{diag}\{e^{ - \lambda h_1^2 / 2}, ..., e^{ - \lambda h_N^2 / 2}\}_{x,y}\,.
\end{equation}
Then, the determinant of $(B^2)_{x,y}$ is substituted by
\begin{equation}
e^{ - \lambda / 2 \sum_x  h_x^2} \det (B^2)_{x,y} = \det (T^2 B^2)_{x,y} = \det (T B B^T T)_{x,y} \equiv \det (Q^2)_{x,y}\,.
\end{equation}
The matrix $Q_{x,y} = (T B)_{x,y}$ is bounded. We normalize it as $\widetilde{Q}_{x,y} \equiv Q_{x,y} / Q_{max}(\lambda)$, 
where $Q_{max}(\lambda)$ is the largest eigenvalue
of $Q_{x,y}$. The determinant $\det (\widetilde{Q}^2)_{x,y} \in (0;1]$ and can be inverted with $n$ flavors of bosonic fields
\begin{eqnarray}
\nonumber
\det (Q^2)_{x,y} &=&  Q_{max}^{2N}(\lambda) \det (\widetilde{Q}^2)_{x,y} \\ 
\nonumber
&=& Q_{max}^{2N}(\lambda)  \lim_{n \rightarrow \infty} \prod_{k = 1..n;\newline x = 1..N}\int [d\phi_{k, x}] [d\phi^\dagger_{k, x}]\\ 
&\,&e^{-\sum_x \sum_{k = 1}^{n}\big[|\sum_y [(\widetilde{Q}_{x,y} - \mu_k \delta_{x,y}) \phi_{k, y}]|^2 +\nu_k^2 |\phi_{k, x}|^2\big]}\,,
\end{eqnarray}
where $\mu_k, \nu_k$ are real constants \cite{Luscher1994}. 
The partition function can be expressed as
\begin{eqnarray}
\nonumber
{\cal Z} = \frac{1}{(2\pi)^N}\lim_{\lambda \rightarrow 0} Q_{max}^{2N}(\lambda) \lim_{n \rightarrow \infty} \int [d\phi_{k, x}] [d\phi^\dagger_{k, x}] \int [dU_x]\\
\nonumber
\int_{-\infty}^{\infty} [df_x] \int_{-\infty}^{\infty} [dh_x]\,
e^{-S_G +  \sum_x\big[ f_{x} + i h_x f_{x} - \lambda (h_x^2 + f_x^2) \big]}\\
e^{-\sum_x \sum_{k = 1}^{n}\big[|\sum_y [[e^{ - \lambda h_x^2 / 2} (h_x H_{x, y} - i A_{x, y}) / Q_{max}(\lambda) - \mu_k \delta_{x,y}] \phi_{k, y}]|^2 +\nu_k^2 |\phi_{k, x}|^2\big]}\,.
\label{LushZ}
\end{eqnarray}
As in the previous section, the limits in (\ref{LushZ}) are not interchangeable.

\section{Conclusions}
The presented hermitization procedure is applicable to any fermionic theory, which may be reduced
to the fermion determinant, i.e. theories with an action that is bilinear in $\psi$ and $\bar\psi$. 
Therefore, we have constructed a bosonization of complex fermionic actions with even number of flavors
and proved the fundamental relation between the $d$-dimensional fermions and $(d+1)$-dimensional bosons.

The suggested bosonization is highly interesting also from a computational point of view. 
One of the promising approaches to calculations on the lattice and path integrals is the
convergent perturbation theory \cite{Belokurov1, Belokurov2, Belokurov3, BelSolSha97, BelSolSha99}.
It was initially formulated only for purely bosonic models and
the extension to theories containing fermions can be now obtained
using bosonization. For lattice QED with an even number of flavors
at zero chemical potential this was done in \cite{Sazonov2014}.
The current work extends the applicability of the convergent perturbation theory and opens a new approach to
lattice theories with a complex action problem.

\section{Acknowledgments}
I acknowledge A. Alexandrova and C. Gattringer for discussions and for reading the manuscript.
This work was supported by the Austrian Science Fund FWF Grant Nr. I 1452-N27.

\bibliography{bosonbib}{}
\bibliographystyle{unsrt}

\end{document}